\documentclass[12pt,twoside]{article}
\usepackage{fleqn,espcrc1}
\usepackage{amssymb,amsbsy}
\usepackage{graphicx}

\def\raw{\rightarrow}

\def\be{\begin{equation}}
\def\ee{\end{equation}}
\def\bea{\begin{eqnarray}}
\def\eea{\end{eqnarray}}
\def\bear{\begin{array}}
\def\ear{\end{array}}
\def\bfig{\begin{figure}}
\def\efig{\end{figure}}
\def\bcen{\begin{center}}
\def\ecen{\end{center}}

\def\Jl#1#2#3#4{{#1} {#2} (#3) #4}
\def\NPA{Nucl. Phys. A}
\def\NPB{Nucl. Phys. B}
\def\PLB{Phys. Lett. B}
 
\def\PREP{Phys. Rep.} 
\def\PRC{Phys. Rev. C}
\def\PRD{Phys. Rev. D}

\title{Two-pion decay modes of the $N^*(1440)$ in
$n p \rightarrow d \pi \pi$}

\author{L. Alvarez-Ruso 
\address{Istituto Nazionale di Fisica Nucleare, Sezione di Torino \\
via P. Giuria 1, I-10125 Torino, Italy}}

\begin{document}
\maketitle

\begin{abstract}
A simple model for the $n p \raw d \pi \pi$ reaction has been developed. 
It is shown that
the deuteron momentum spectra measured at $T_n = 795$ MeV can be understood
in terms of the Roper excitation and its $N\pi\pi$ decay modes. A similar
pattern, recently observed in $pp \raw p p \pi^+ \pi^-$, can be explained in
the same way. 
\end{abstract}
\vspace{.5cm}

Double-pion production in nucleon-nucleon ($NN$) collisions is a source of interesting information about
the properties of baryonic resonances and the $NN$ interaction in the inelastic region.
It has actually become an active field of experimental research at CELSIUS
and COSY, where the reactions $p p \raw N N \pi \pi, d \pi \pi$, 
$p d \raw {^3}He \, \pi \pi$ and $d d \rightarrow {^4}He \,\pi \pi$ are being
studied near threshold.    

The $n p \raw d \pi \pi$ reaction was studied in the past in
connection with the ABC effect. Here, I focus the attention on
the deuteron spectrum measured using a neutron beam with 
$T_n = 795$~MeV \cite{hollas}. In contrast to the experiments performed at
higher energies ($\sim 1$~GeV), the ABC peaks
are not present in the data; they rather show a well defined bump at high
$\pi \pi$ missing masses, in disagreement with the models available in the
literature \cite{hollas}. 

My aim is to show that this behavior can be understood as a
consequence of the interference of two mechanisms involving the excitation of
the Roper resonance $N^*(1440)$ and its subsequent decay into
$N (\pi \pi)^{T=0}_{S-wave}$ and $\Delta \pi$ respectively \cite{yo2}.
The present model is a reduced
version of the one of Ref.~\cite{yo1} for the $NN \raw NN \pi \pi$ reaction. 
\begin{figure}[b!]
\begin{center}
\includegraphics[height=.7\textwidth, angle=-90]{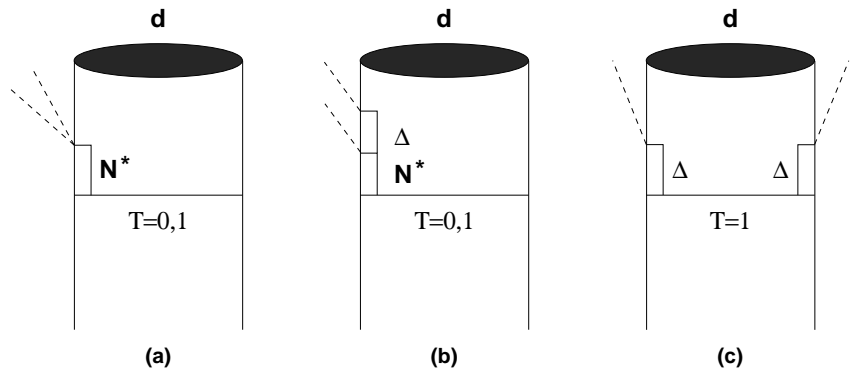}
\vspace{-.5cm}
\caption{Set of diagrams of the model.}
\label{diag2}
\end{center}
\vspace{-.5cm}
\end{figure}
With these ingredients, and using the Paris deuteron wave function, 
one can calculate the deuteron momentum distribution for different laboratory angles.
\begin{figure}[t!]
\begin{center}
\includegraphics[height=\textwidth, angle=-90]{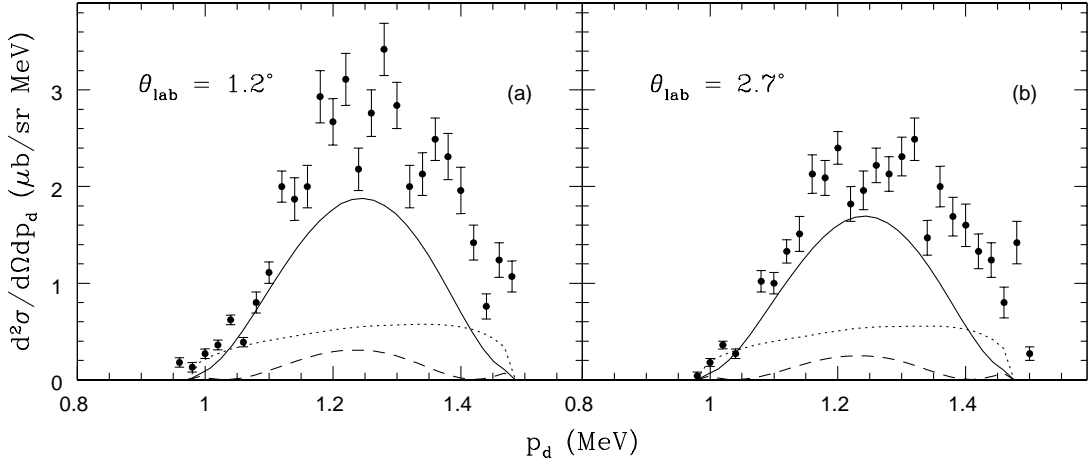}
\vspace{-1cm}
\caption{Calculated deuteron momentum spectra 
at $T_n = 795$~MeV/c (solid lines) compared
to the measured data \cite{hollas}. The dotted line
corresponds to the $N^* \rightarrow N (\pi \pi)^{T=0}_{S-wave}$ mechanism
(Fig.~\ref{diag2}~a); the short-dashed line stands for the
$N^* \rightarrow \Delta \pi$  (Fig.~\ref{diag2}~b).}
\label{all}
\end{center}
\vspace{-.7cm}
\end{figure}
The mechanism $N^* \rightarrow N (\pi \pi)^{T=0}_{S-wave}$ 
produces spectra very similar to phase space, while the $N^* \rightarrow \Delta \pi$ 
mechanism plays a crucial role in providing
the right shape to the distributions through its interference with the larger
$N^* \rightarrow N (\pi \pi)^{T=0}_{S-wave}$ contribution. This interference is 
constructive at high $\pi \pi$ masses and destructive at low ones.  
Such a pattern can be
understood by realizing that the $N^* \rightarrow \Delta \pi$ amplitude is
dominated by terms proportional to the scalar product of the outgoing pions
three momenta; this scalar product has different signs in the center of the
spectra, where the pions go back to back, and at the edges, where they
travel together. 
\begin{figure}[b!]
\begin{minipage}{.45\linewidth}
\begin{center}
\vspace{-1.3cm}
\includegraphics[height=\textwidth,angle=-90]{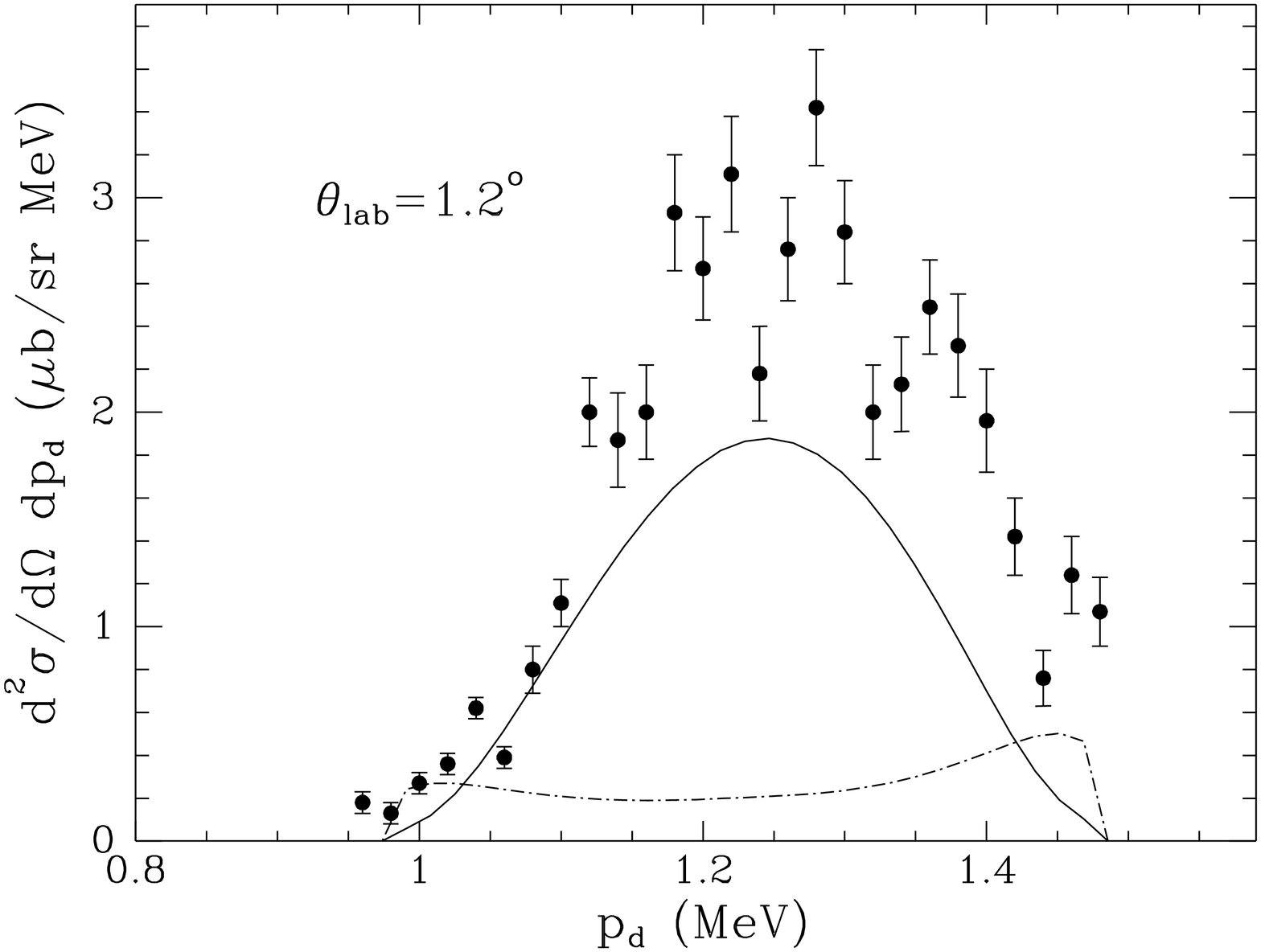}
\vspace{-.7cm}
\caption{Calculated spectra for two different
choices of the $g_{N^* \Delta \pi}$ sign}
\label{signo}
\end{center}
\end{minipage}
\hfill
\begin{minipage}{.45\linewidth}
\begin{center}
\includegraphics[height=\textwidth,angle=-90]{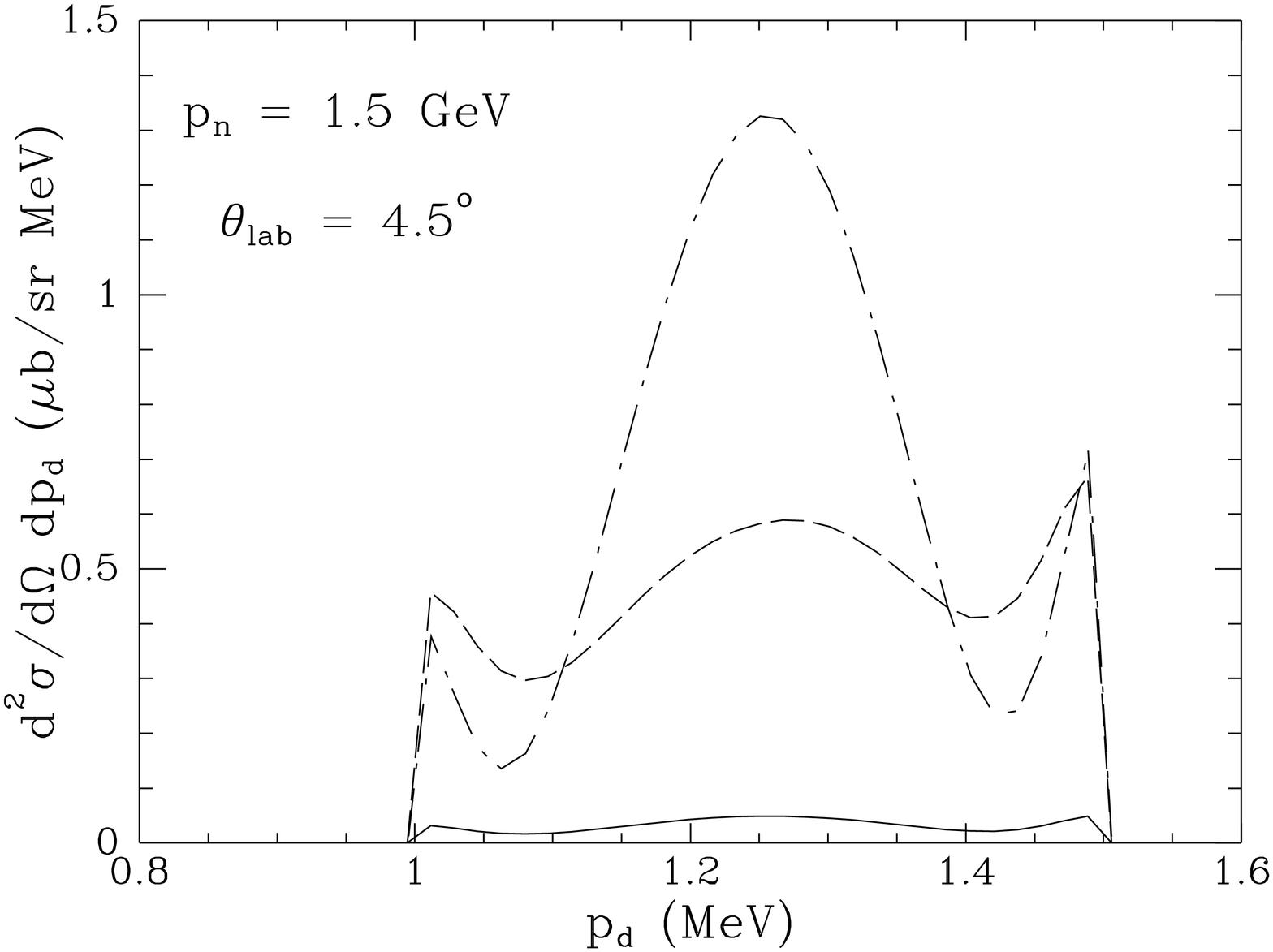}
\vspace{-.6cm}
\caption{The double-$\Delta$ mechanism in the case of $\pi$
exchange alone (dashed line), $\pi + \rho$ (dash-dotted) and
$\pi + \rho + $ short range correlations (solid line).}
\label{deldel}
\end{center}
\end{minipage}
\vspace{-.3cm}
\end{figure}
In order to further illustrate the effect of the interference,
Fig.~\ref{signo} shows the effect of changing the relative sign of the two
amplitudes. The data clearly favor a choice of the sign of 
$g_{N^* \Delta \pi}$ in agreement with earlier works \cite{manolo}.

The double-$\Delta$ mechanism is too small to be
represented in Fig. \ref{all}. In Fig.~\ref{deldel}, its contribution
is shown for $T_n=830$~MeV, 
$\theta_{lab}=4.5^{\circ}$
and using the Hulthen wave function, in order to compare with the calculation 
of Ref.~\cite{bnrs} (Fig.~4~b). The differential
cross section obtained in the case of only pion exchange  is very similar
to the one given by the relativistic model of  Ref.~\cite{bnrs} but the
inclusion of the rho exchange modifies the result, and the short range
correlations cause a strong reduction of the strength of this mechanism.

The presented spectra are equivalent to a $\pi^+ \pi^-$
invariant mass distribution shifted to higher masses with respect
to phase space; a similar behavior has been recently observed in  
$pp \raw p p \pi^+ \pi^-$ at a lower energy $T_p=750$~MeV 
(see Fig. 2 of Ref.~\cite{wasa}). I have calculated this
observable including only the dominant isoscalar excitation of the $N^*(1440)$
decaying into $N (\pi \pi)^{T=0}_{S-wave}$ and $\Delta \pi$. I have also
included $pp$ final state interaction (strong plus Coulomb) via effective
range approximation \cite{weyer} and the experimental kinematical
constrain $3^\circ \leq \theta_{lab} \leq 24^\circ$ for the
detected particles (protons and $\pi^+$). Also here,  
the interference between the two mechanisms allows to explain the observed
shift.  
\begin{figure}[h!]
\vspace{-.5cm}
\bcen
\includegraphics[height=0.45\textwidth]{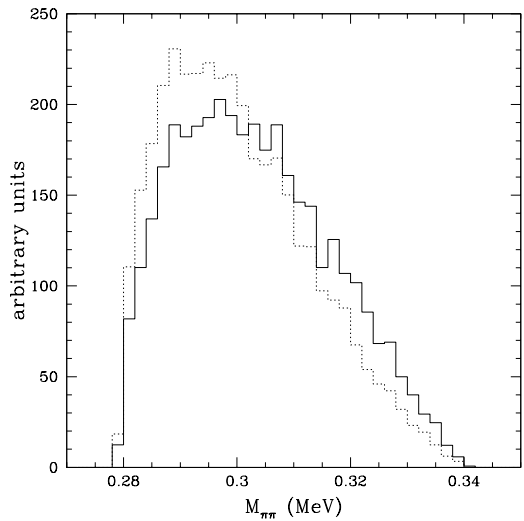}
\vspace{-.8cm}
\caption{$\pi^+ \pi^-$ invariant mass spectrum for $pp \raw p p \pi^+ \pi^-$ at 
$T_p=750$~MeV. The dotted line stands for 
$N^* \rightarrow N (\pi \pi)^{T=0}_{S-wave}$ alone; the solid line
includes also $N^* \rightarrow \Delta \pi$.}
\label{minv}
\end{center}
\vspace{-1cm}
\end{figure}



\begin{thebibliography}{9}
\bibitem{hollas} C. L. Hollas et al., 
\Jl{\PRC}{25}{1982}{2614}.

\bibitem{yo2} L. Alvarez-Ruso,
\Jl{\PLB}{452}{1999}{207}.

\bibitem{yo1} L. Alvarez-Ruso, E. Oset, E. Hern\'andez,
\Jl{\NPA}{633}{1998}{519}.

\bibitem{manolo}  D. M. Manley, E. M. Saleski, 
\Jl{\PRD}{45}{1992}{4002};
V. Sossi et al., \Jl{\NPA}{548}{1992}{562}.

\bibitem{bnrs} I. Bar-Nir, T. Risser, M. D. Shuster, 
\Jl{\NPB}{87}{1975}{109}.

\bibitem{wasa} R. Bilger et al., \Jl{\NPA}{663-664}{2000}{469c}.

\bibitem{weyer} H. J. Weyer, \Jl{\PREP}{195}{1990}{295}. 
\end{thebibliography}
\end{document}